\input harvmac
\input psfig.sty
\input labeldefs.tmp
\writedefs

\def\TR{$\displaystyle{{}##}$\hfil}
\def\TC{\hfil$\displaystyle{##}$\hfil}
\def\TT{\hbox{##}}
\def\seqalign#1#2{\vcenter{\openup1\jot
  \halign{\strut #1\cr #2 \cr}}}

\def\comment#1{}
\def\fixit#1{}

\def\tf#1#2{{\textstyle{#1 \over #2}}}

\def\mop#1{\mathop{\rm #1}\nolimits}

\def\sqr#1#2{{\vcenter{\vbox{\hrule height.#2pt
       \hbox{\vrule width.#2pt height#1pt \kern#1pt
            \vrule width.#2pt}
         \hrule height.#2pt}}}}
\def\square{\mathop{\mathchoice\sqr56\sqr56\sqr{3.75}4\sqr34\,}\nolimits}

\def\litem#1#2{{\leftskip=20pt\noindent\hskip-10pt {\bf #1} #2\par}}

\def\diag#1{{\rm diag}\left\{#1\right\}}

\def\href#1#2{#2}  
\def\em{\it}
\def\IB{\relax\hbox{$\inbar\kern-.3em{\rm B}$}}
\def\IC{\relax\hbox{$\inbar\kern-.3em{\rm C}$}}
\def\ID{\relax\hbox{$\inbar\kern-.3em{\rm D}$}}
\def\IE{\relax\hbox{$\inbar\kern-.3em{\rm E}$}}
\def\IF{\relax\hbox{$\inbar\kern-.3em{\rm F}$}}
\def\IG{\relax\hbox{$\inbar\kern-.3em{\rm G}$}}
\def\IGa{\relax\hbox{${\rm I}\kern-.18em\Gamma$}}
\def\IH{\relax{\rm I\kern-.18em H}}
\def\IK{\relax{\rm I\kern-.18em K}}
\def\IL{\relax{\rm I\kern-.18em L}}
\def\IP{\relax{\rm I\kern-.18em P}}
\def\IR{\relax{\rm I\kern-.18em R}}
\def\IZ{\relax\ifmmode\mathchoice
{\hbox{\cmss Z\kern-.4em Z}}{\hbox{\cmss Z\kern-.4em Z}}
{\lower.9pt\hbox{\cmsss Z\kern-.4em Z}}
{\lower1.2pt\hbox{\cmsss Z\kern-.4em Z}}\else{\cmss Z\kern-.4em
Z}\fi}

\def\II{\relax{\rm I\kern-.18em I}}


\def\CB {{\cal B}}

\def\CN {{\cal N}}

\def\CX {{\cal X}}
\def\CY {{\cal Y}}
\def\CZ {{\cal Z}}

\def\p{\partial}
\def\pb{\bar{\partial}}

\def\jb{{\bar j}}


\def\Tr{\mop{Tr}}
\def\Id{{\rm Id}}
\def\Vol{\mop{Vol}}

\def\p{\partial}
\def\pb{\bar{\partial}}

\def\inbar{\,\vrule height1.5ex width.4pt depth0pt}
\font\cmss=cmss10 \font\cmsss=cmss10 at 7pt

\def\a{\alpha}

\def\b{\beta}

\def\p{\partial}

\def\Z{\relax\ifmmode\mathchoice
{\hbox{\cmss Z\kern-.4em Z}}{\hbox{\cmss Z\kern-.4em Z}}
{\lower.9pt\hbox{\cmsss Z\kern-.4em Z}}
{\lower1.2pt\hbox{\cmsss Z\kern-.4em Z}}\else{\cmss Z\kern-.4em Z}\fi}

\def\r{\bf r}
\font\cmss=cmss10 \font\cmsss=cmss10 at 7pt

\def\CX{{\cal X}}

\lref\gukov{S.~Gukov, ``Comments on $\CN=2$ $AdS$
 Orbifolds'',
hep-th/9806180.}
\lref\witkleb{
I.~R. Klebanov and E.~Witten, ``Superconformal field theory on three-branes at
  a Calabi-Yau singularity,''
  \href{http://xxx.lanl.gov/abs/hep-th/9807080}{{\tt hep-th/9807080}}.}
\lref\dh{P.~Kronheimer, J. Diff. Geom. {\bf 28} (1989) 665, {\it
ibid.} {\bf 29}
(1989) 685\semi
J.~Duistermaat, G.~Heckman, Invent. Math. {\bf 69} (1982) 259.}
\lref\dw{R.~Donagi, E.~Witten, ``Supersymmetric Yang-Mills Systems And
Integrable Systems'', hep-th/9510101,  Nucl.Phys. {\bf B}460 (1996) 299.}
\lref\dm{M.~Douglas, G.~Moore, ``D-Branes, Quivers and ALE
Instantons'',
hep-th/9603167.}
\lref\enh{M.~Douglas, ``Enhanced Gauge Symmetry in M(atrix) Theory'', JHEP 9707
(1997) 004, hep-th/9612126.}
\lref\gdm{M.~Douglas, B.~Greene, D.~Morrison, ``Orbifold resolution by
D-Branes'', hep-th/9704151.}
\lref\uranga{A.~Uranga, ``Brane Configurations for Branes at
Conifolds'', hep-th/9811004.}
\lref\mukhi{K.~Dasgupta, S.~Mukhi, ``Brane Constructions, Conifolds
and M-Theory'', hep-th/9811139.}

\lref\gkBaryon{S.~S. Gubser and I.~R. Klebanov, ``Baryons and domain
walls in an N=1 superconformal gauge theory,'' {\tt hep-th/9808075}.}

\lref\ddg{D.-E. Diaconescu, M.~R. Douglas, and J.~Gomis, ``Fractional
branes and wrapped branes,'' JHEP 02:013 (1998), {\tt
hep-th/9712230}.}

\lref\jpTASI{J.~Polchinski, ``TASI lectures on D-branes,'' {\tt
hep-th/9611050}.}

\lref\malda{
J.~Maldacena, ``The Large N limit of superconformal field theories and
  supergravity,'' {\em Adv. Theor. Math. Phys.} {\bf 2} (1998) 231,
  \href{http://xxx.lanl.gov/abs/hep-th/9711200}{{\tt hep-th/9711200}}.}

\lref\kasil{
S.~Kachru and E.~Silverstein, ``4-D conformal theories and strings on
  orbifolds,'' {\em Phys. Rev. Lett.} {\bf 80} (1998) 4855,
  \href{http://xxx.lanl.gov/abs/hep-th/9802183}{{\tt hep-th/9802183}}.}

\lref\lnv{
A.~Lawrence, N.~Nekrasov, and C.~Vafa, ``On conformal field theories in
  four-dimensions,'' \href{http://xxx.lanl.gov/abs/hep-th/9803015}{{\tt
  hep-th/9803015}}.}

\lref\gEin{
S.~S. Gubser, ``Einstein manifolds and conformal field theories,''
  \href{http://xxx.lanl.gov/abs/hep-th/9807164}{{\tt hep-th/9807164}}.}

\lref\gkp{
S.~S. Gubser, I.~R. Klebanov, and A.~M. Polyakov, ``Gauge theory correlators
  from noncritical string theory,'' {\em Phys. Lett.} {\bf B428} (1998) 105,
  \href{http://xxx.lanl.gov/abs/hep-th/9802109}{{\tt hep-th/9802109}}.}

\lref\witadsi{
E.~Witten, ``Anti-de Sitter space and holography,'' {\em Adv. Theor. Math.
  Phys.} {\bf 2} (1998) 253, \href{http://xxx.lanl.gov/abs/hep-th/9802150}{{\tt
  hep-th/9802150}}.}

\lref\LeighStrassler{
 R.~G. Leigh and M.~J. Strassler, ``Exactly marginal operators and duality
in four-dimensional ${\cal N}=1$ supersymmetric gauge theory,''
{\em Nucl. Phys.} {\bf B}447 (1995) 95, {\tt hep-th/9503121}.}

\lref\kls{ A.~Karch, D.~Lust, D.~J.~Smith, ``Equivalence of Geometric
      Engineering and Hanany-Witten via Fractional Branes'',
      hep-th/9803232,
Nucl.Phys. {\bf B} 533 (1998) 348-372}

\lref\witadsii{
E.~Witten, ``Anti-de Sitter space, thermal phase transition, and confinement in
  gauge theories,'' {\em Adv. Theor. Math. Phys.} {\bf 2} (1998) 505,
  \href{http://xxx.lanl.gov/abs/hep-th/9803131}{{\tt hep-th/9803131}}.}

\lref\girych{
G.~W. Gibbons and P.~Rychenkova, 
``HyperK\"{a}hler Quotient Construction of BPS Monopole Moduli Spaces''
  \href{http://xxx.lanl.gov/abs/hep-th/9608085}{{\tt hep-th/9608085}}.}

\lref\arnold{V.I.~Arnold, S.~Gusein-Zade, A.~Varchenko, ``Singularities of
  Differentiable Maps'', Vol. I, Monographs in Mathematics, 
{\rm vol. 82}, Birkh\"auser, Boston, Basel, Stuttgart, 1985.}

\lref\MorPless{
D.~R. Morrison and M.~R. Plesser, ``Nonspherical horizons. 1,''
  \href{http://xxx.lanl.gov/abs/hep-th/9810201}{{\tt hep-th/9810201}}.}

\lref\romans{
L.~J. Romans, ``New compactifications of chiral N=2 d=10 supergravity,'' {\em
  Phys. Lett.} {\bf 153B} (1985) 392.}

\lref\kehag{
A.~Kehagias, ``New type IIB vacua and their F theory interpretation,'' {\em
  Phys. Lett.} {\bf B435} (1998) 337,
  \href{http://xxx.lanl.gov/abs/hep-th/9805131}{{\tt hep-th/9805131}}.}

\lref\GukKap{
S.~Gukov and A.~Kapustin, ``New N=2 superconformal field theories from M / F
  theory orbifolds,'' \href{http://xxx.lanl.gov/abs/hep-th/9808175}{{\tt
  hep-th/9808175}}.}

\lref\grw{
M.~Gunaydin, L.~J. Romans, and N.~P. Warner, ``Compact and noncompact gauged
  supergravity in five-dimensions,'' {\em Nucl. Phys.} {\bf B272} (1986) 598.}

\lref\ppz{
L.~Girardello, M.~Petrini, M.~Porrati, and A.~Zaffaroni, ``Novel local CFT and
  exact results on perturbations of N=4 superYang Mills from AdS dynamics,''
  \href{http://xxx.lanl.gov/abs/hep-th/9810126}{{\tt hep-th/9810126}}.}

\lref\dz{
J.~Distler and F.~Zamora, ``Nonsupersymmetric conformal field theories from
  stable anti-de Sitter spaces,''
  \href{http://xxx.lanl.gov/abs/hep-th/9810206}{{\tt hep-th/9810206}}.}

\Title{ \vbox{\baselineskip12pt\hbox{hep-th/9811230}\hbox{HUTP- 98/A051}
\hbox{YCTP-P29-98}
\hbox{ITEP-TH-64/98}
}}
{\vbox{\centerline{Generalized Conifolds and}
\centerline{Four Dimensional }
\centerline{$\CN=1$  Superconformal Theories}
}}

\centerline{Steven S. Gubser$^1$, Nikita Nekrasov$^2$ 
and Samson Shatashvili\footnote{$^{3}$}{On leave of absence from St. Petersburg 
Steklov Mathematical
Institute}}

{\it $^2$ Institute
 of Theoretical and Experimental
Physics, 117259, Moscow, Russia}

{\it $^{1,2}$ Lyman Laboratory of Physics,
Harvard University, Cambridge, MA 02138, USA}

{\it $^{3}$ CERN, Theory Division, 1211 Geneva 23, Switzerland}

{\it $^{3}$ Department of Physics, Yale University, New Haven, CT
06520 USA}

\vskip .1in
\centerline{ssgubser@bohr.harvard.edu, nikita@string.harvard.edu,
shatash@mail.cern.ch} 
\vskip .3in

{ 
This paper lays groundwork for the detailed study of the non-trivial
renormalization
group flow connecting supersymmetric fixed points in four dimensions
using string theory on AdS spaces. Specifically, 
we consider D3-branes placed at singularities of Calabi-Yau
threefolds which generalize the conifold singularity and have an ADE
classification.  The $\CN=1$ superconformal theories dictating their
low-energy dynamics are infrared fixed points arising from
deforming the corresponding ADE $\CN=2$ superconformal field theories
by mass terms for adjoint chiral fields.  We probe the geometry with a
single $D3$-brane and discuss the 
near-horizon supergravity solution for a 
large number $N$ of coincident $D3$-branes.
 }

\bigskip

\Date{11/98}

\baselineskip14pt
\newsec{Introduction}

The recently proposed duality \malda\gkp\witadsi\ between string
theory on a space $B$ of constant negative curvature and certain gauge
theories which live on the boundary of $B$ provides fascinating
possibilities for the study of both sides of the equivalence.  The
original conjecture \malda\ identifies type~IIB string theory on
$AdS_{5} \times S^{5}$ with four-dimensional $\CN=4$ super-Yang-Mills
theory with gauge group $SU(N)$.  In gauge theory terms, the validity
of the supergravity approximation to type~IIB string theory depends on
having both $N$ and the 't~Hooft coupling $g_{YM}^2 N$ large.

The conjecture has been extended \kasil\ to the spaces of the form
$AdS_{5} \times X^{5}$, where $X^{5} = S^{5}/{\Gamma}$, with $\Gamma$
being a discrete subgroup of $SO(6)$.  The corresponding gauge
theories have been described in \lnv.  They have $\CN=2$, $1$, or $0$
superconformal symmetry according as $\Gamma$ is a subgroup of
$SU(2)$, $SU(3)$, or $SU(4) \approx SO(6)$.  The low-energy dynamics
of $N$ D3-branes placed at an orbifold singularity of a Calabi-Yau
three-fold is described by one of these gauge theories.

In general one could consider string theory on $AdS_5 \times M_5$
where $M_5$ is an arbitrary Einstein manifold.  This Freund-Rubin
ansatz is the most general static bosonic near-horizon geometry with
only the metric and the self-dual five-form excited.  Dimension five
is the first where there are infinitely many different Einstein
manifolds which are not even locally diffeomorphic, and a natural
question to ask is what all the corresponding field theories are.  The
D-brane origin of the holographic conjecture suggests a two step
approach to finding the answer: first find a manifold with an isolated
singularity such that the near-horizon geometry in supergravity of a
black three-brane located on this singularity is $AdS_5 \times M_5$;
then figure out the field theory of D3-branes moving close to that
singularity.  In practice, we may start with a known singularity, work
out from supergravity the near-horizon geometry of black three-branes
on the singularity, construct a gauge theory describing D3-branes near
the singularity, and consider the result as a holographic dual pair.
As a rule, the gauge theory is worked out in the approximation that the
D3-branes do not significantly distort the geometry.  This
approximation is correct in the limit of weak coupling, whereas
supergravity is valid at strong coupling.  If the gauge theory is
superconformal, we may feel confident in extrapolating it to strong
coupling so that the comparison with supergravity can be made
directly.  The ``extrapolation'' of supergravity down to weak coupling
is much harder because it requires the full type~IIB string theory in
a background with Ramond-Ramond fields.

The first successful example of this approach for a manifold not
locally diffeomorphic to $S^5$ was \witkleb.  There the space $M_5 =
T^{1,1} \approx \left( SU(2) \times SU(2) \right)/U(1)$ was
considered, which is the base of what we will call the $A_1$ conifold:
  \eqn\StandardConifold{
   X^{2} + Y^{2} + Z^{2} + T^{2} = 0
  }
 $N$ $D3$-branes which are placed at the conifold singularity are
described by a $\CN=1$ superconformal field theory which is a
non-trivial infrared fixed point of the renormalization group.
While this work was in progress, a further class of examples was
worked out in \MorPless\ using toric geometry.

The purpose of this paper is to construct holographic dual pairs out
of an infinite class of conical singularities.  The geometry
\StandardConifold\ is a fibration of a four-dimensional ALE space of
type $A_1$ over the complex plane; our singular geometries will be
fibrations of ALE spaces of arbitrary $ADE$ type, and we will call
them $ADE$ conifolds.  The field theory constructed in \witkleb\
descends by RG flow from the $\CN = 2$ ${\bf S}^5/\IZ_2$ orbifold
theory with mass terms for chiral fields added to break the
supersymmetry to $\CN = 1$; our field theories descend from mass
deformations of the general $A,D,E$ type 
$\CN = 2$ orbifold
theories.  Unlike the $A_{1}$ case there is a moduli space of such
mass deformations which is isomorphic to the projectivization of the
moduli space of the versal deformation of the corresponding
singularity.  In all cases we will have $\CN = 1$ superconformal
symmetry, which is one quarter of maximal (eight real supercharges).
Our $ADE$ conifold geometries are non-compact, but they can all be
realized as singularities of compact Calabi-Yau three-folds.  Our
results are most complete for the $A_k$ conifolds, but on many points
we include also the analysis for the $D_k$ and $E_k$ cases.

Section~\GaugeTheory\ is devoted to the study of D3-branes near the
orbifold singularities from which our conifold theories descend.  If
the D3-branes moving in a given singular geometry are claimed to be
described by the infrared limit of a particular gauge theory, then the
first thing that should be verified is that this gauge theory
specialized to a single D3-brane has for its moduli space precisely
the singular geometry in question.  We present a formal argument why
this should be so for the ADE conifold singularities.  For $\Gamma =
A_k$ or $D_k$, we present an explicit construction of the Higgs
branch, $\IC^2/\Gamma$, in the case where the orbifold is not
deformed.  In the case where it is deformed, we show how the
deformation parameters are related to the periods of the complex
two-form.  For $\Gamma = A_k$, we show explicitly how the conifold
arises from the solution of the F- and D-flatness conditions; for the
$D_k$ and $E_k$ cases we fall back on the formal argument presented
earlier.  Finally, we calculate for $\Gamma=A_k$ the Kahler metric
from gauge theory at the classical level, exhibit its cone structure
and
 observe that it is not the
Calabi-Yau metric. We then briefly discuss the reasons for that,
in agreement with the results  of \gdm.

In section~\Supergravity\ we outline the supergravity side of the dual
pair.  Writing out explicit metrics for the Einstein spaces seems
impossible since Calabi-Yau metrics are not known in closed form for
the general $ADE$ conifolds.  However, we exploit a natural action of
$\IC^*$ on the conifold geometry to show that the spectrum of chiral
primary operators in the gauge theory is correctly reproduced by the
holographic mass-dimension relation.
We then proceed with more detailed analysis of the blowup modes
and the corresponding AdS supergravity multiplets. 

In section~\DualBranes\ we present the realization (in the $A_{k}$
case)
of our gauge theories by using other branes as a background instead of
the singular geometry. 

In the section~\Conclusions\ we present our conclusions and some
conjectures.

\newsec{Gauge theory perspective}
\seclab\GaugeTheory

In this section we describe the geometry of ALE spaces and present
a construction
of the CY threefold obtained by fibering ALE space over a one-dimensional
base. This construction is obtained by looking at the Higgs branch of
the gauge theory on the world-volume of a single $D3$-brane. 

\subsec{Single $D$-brane on ALE space}
\subseclab\FirstConstruct

  \topinsert{\centerline{\psfig{figure=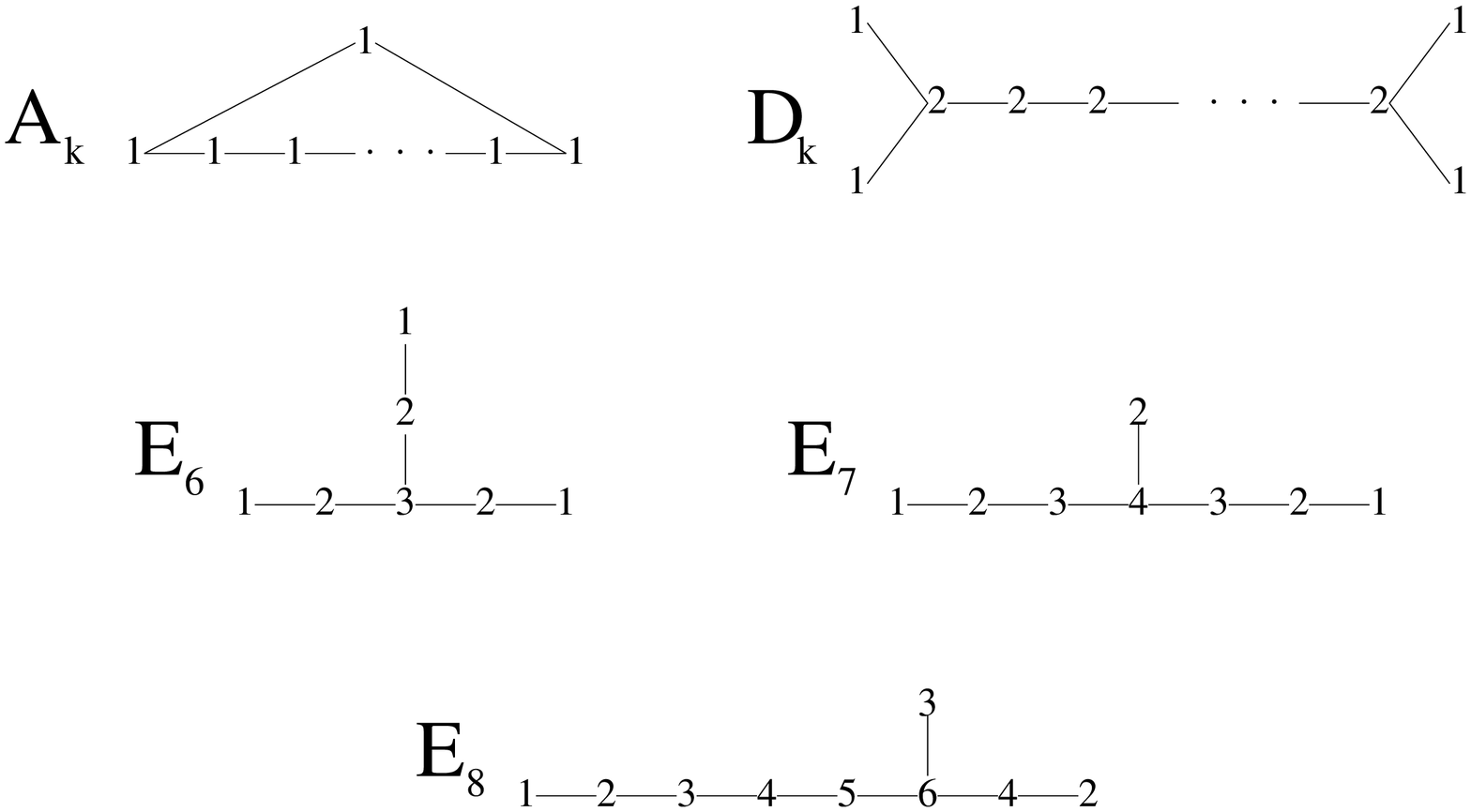,width=3in}}
   {\medskip\narrower\narrower\noindent\baselineskip11pt
    Figure 1: The extended Dynkin diagrams of ADE type, including the 
    indices $n_i$ of each vertex.
   }}\endinsert 
First of all we recall the construction of the gauge theory
on the world-volume of a single $D$-brane placed at the
orbifold singularity $\IC^{2}/{\Gamma}$, where $\Gamma$
is a discrete subgroup of $SU(2)$ of ADE type. 
The field theory has $\CN=2$ supersymmetry. Its gauge group is the
product:
\eqn\ggr{G_{1} = \times_{i=0}^{r} U(n_{i})}
where $i$ runs through the set of vertices of
extended Dynkin diagram of the corresponding ADE type (see figure~1),
 or, equivalently, through the set of irreducible representations
$\r_{i}$ of $\Gamma$. The label $i=0$ corresponds to the trivial
representation. The number $n_{i}$ is simply the dimension of
$\r_{i}$.  Let $h = \sum_i n_i$. This number coincides with the
dual Coxeter number of the corresponding ADE Lie group.

The matter content of our gauge theory is that of $a_{ij}$ bi-fundamental
hypermultplets in the representations $(n_{i}, {\bar n}_{j})$, where
$a_{ij}$ is determined from the decomposition: 

\eqn\dcmps{{\IC}^{2}
\otimes {\r}_{i} = \oplus_{j} {\IC}^{a_{ij}} \r_{j}} 
{}From the $\CN=1$
point of view each pair $i,j$ with $a_{ij} \neq 0$ gives rise to 
a pair of chiral multiplets, call them $(B_{ij}, B_{ji})$. 
$B_{ij}$ is a complex matrix, transforming in the $(n_i, \bar n_{j})$ of the 
$i$'th and $j$'th gauge groups. 
The theory has a superpotential: 
  \eqn\sprptn{W =
   \sum_{i} {\Tr} \mu_{i} \phi_{i}} 
 where $\phi_{i}$ is the scalar of the
$i$'th vector multiplet and $\mu_{i}$ is the complex moment map.
There is some gauge freedom in the choice of explicit expressions for
$\mu_i$.  Let us introduce an antisymmetric adjacency matrix $s_{ij}$
for the extended Dynkin diagram, such that $s_{ij} = \pm 1$ when $i$
and $j$ are adjacent nodes and the sign, which is part of our gauge
choice, indicates a direction on the edge between them.  Then we can
write
 \eqn\ChoseMu{\mu_i{}^{\alpha_i}{}_{\beta_i} = 
    \sum_j s_{ij} B_{ij}{}^{\alpha_i}{}_{\gamma_j} 
B_{ji}{}^{\gamma_j}{}_{\beta_i} 
\ .}
 Here an upper index $\alpha_i$ indicates a fundamental representation
of $U(n_i)$, while a lower index $\alpha_j$ indicates an
anti-fundamental representation of $U(n_i)$.  We will suppress these
color indices when their contractions are clear from context.  There
is one relation among the $\mu_i$:
\eqn\TraceMu{
   \sum_{i} {\Tr}\mu_{i} = 0 \ .}

Without breaking $\CN=2$ supersymmetry one may introduce complex FI terms
which modify superpotential as follows:
$$
W \to W - \sum_{i} \zeta_{i} {\Tr} \phi_{i}
$$ 
The moduli space of vacua (Higgs branch) of the theory with
$\zeta_{i}$'s coincides with the complex deformation of
the orbifold $\IC^{2}/\Gamma$ into the (smooth for generic $\zeta_i$'s)
ALE space $S_{\zeta}$. 
It can be described as a quotient of the space of solutions of the
equations
\eqn\cmpl{\mu_{i} = \zeta_{i} \Id_{n_{i}} }
by the complexification of the gauge group 
$$
G_{1}^{c} = \times_{i} {\rm GL}_{n_{i}}({\IC})
$$
For $\zeta = 0$
the Higgs branch becomes the singular orbifold, and at the singularity
the Coulomb branch appears. At the generic point of the Higgs branch there
is one massless vector multiplet, which 
corresponds to the $U(1)$ subgroup of $G_{1}$,
which is embedded diagonally into each $U(n_{i})$. 
Since all the matter is in the bi-fundamentals, it is neutral
with respect to this $U(1)$ subgroup.

It is important to notice that the holomorphic
$2$-form $\omega_{\zeta}$ of $S_{\zeta}$ has periods which depend
linearly on $\zeta$ (it follows from the complexification
of Duistermaat-Heckmann theorem \dh\dw). This observation will be used below.

We would like to list here the equations which describe $S_{0}$
for various $\Gamma$ as hypersurfaces $f_{\Gamma} = 0$ in the space
 $\IC^{3}$ with
coordinates $x, y, z$:
\eqn\equ{\eqalign{A_{k}: \quad & f_{\Gamma} = x^{k+1} + y^{2} + z^{2}\cr
D_{k} : \quad & f_{\Gamma} = x^{k-1} + x y^{2} + z^{2} \cr
E_{6}: \quad & f_{\Gamma} = x^{4} + y^{3} + z^{2}\cr
E_{7} : \quad & f_{\Gamma} = x^{3}y +  y^{3} + z^{2}\cr
E_{8}: \quad & f_{\Gamma} = x^{5} + y^{3} + z^{2}\cr}}
The  equations $f_{\Gamma} = 0$ are invariant under a $\IC^*$ action, which
is specified by giving 
the weights $\alpha, \beta$ to the coordinates $x, y$ and $z$ as follows:
\eqn\wght{\seqalign{\span\TR\ \ & \span\TC\ \ & \span\TC\ \ & \span\TC\ \ & 
  \span\TC}{
\Gamma & \alpha = [x] & \beta = [y] & [z] = h/2 & h 
  \cr\noalign{\vskip1.5\jot}
A_{k} & 1 & {\textstyle{ {k+1}\over{2}}} & {\textstyle{ {k+1}\over{2}}} & 
  k+1\cr
D_{k} & 2 & k - 2& k-1 & 2(k-1)\cr
E_{6}& 3 & 4 & 6 & 12\cr
E_{7}& 4 & 6 & 9 & 18\cr
E_{8} & 6 & 10 & 15 & 30}}
Notice that $\alpha + \beta = 1 + {{h}\over{2}}$.

The complex deformation of the surface $S_{0}$ is described by the equation
$$
f(x,y; t) + z^{2}= 0
$$
where $t$ are some coordinates on the  base of deformation. 
There are canonical formulae, listed, say in \arnold, which represent
$f(x,y;t )$ as polynomials in $x, y$. 

\eqn\eee{\eqalign{
A_{k}: \quad & f_{\Gamma}= P_{k+1}(x) + y^{2} + z^{2}\cr
D_{k}: \quad & f_{\Gamma}= x^{k-1} + Q_{k-2}(x) + t_{0} y + x y^{2} + z^{2} \cr
E_{6}: \quad & f_{\Gamma}= y^{3} + Q_{2}(x) y + P_{4}(x) + z^{2}\cr
E_{7}: \quad & f_{\Gamma}= y^{3} + P_{3}(x) y + Q_{4}(x) + z^{2}\cr
E_{8}: \quad & f_{\Gamma}= y^{3} + Q_{3}(x) y + P_{5}(x) + z^{2}\cr}}
where $P_{k}(x) = x^{k} + \sum_{\ell=1}^{k-1} t_{\ell} x^{k-\ell-1}$, 
$Q_{k} (x) = \sum_{l = 1}^{k+1} t_{l} x^{k-l+1}$. 
We now wish to relate the coordinates $t_{k}$ and $\zeta_{l}$'s. 
In order to do so we study the periods of the holomorphic two-form:
$$
\omega_{\zeta} = {{dx \wedge dy \wedge dz}\over{df_{\Gamma}}}
$$
\litem{{\it $A_{k}$ case.}}{In this case the ALE space is a fibration over the
$x$-plane, whose fiber is isomorphic to $\IC^{*}$ for $x \neq x_i$ where
$x_i$ are the roots of $P_{k+1}(x_i)=0, i= 0, \ldots, k$.
The form $\omega_{\zeta}$ factorizes as:
$\omega_{\zeta} = {{dy}\over{2z}} \wedge dx$. To get a non-trivial
period of it we choose a one-dimensional contour in the $x$-plane which
connects
$x_i$ and $x_j$ for $i \neq j$ and doesn't pass through other
$x_k$'s. The fiber over its generic point contains
a non-trivial one-cycle, over which the form $dy/2z$ integrates to $\pi$ 
(write $y^2 + z^2 = r^2$, $y = r {\rm sin} \alpha$, $ z = r {\rm cos}
\alpha$, $0 \leq \alpha < 2\pi$, 
$r$ is determined by $x$ hence $dy/2z =  {1\over 2} d\alpha$.)
Hence we get
\eqn\prds{\left[ \omega_{\zeta} \right]_{ij} = \pi \left( x_i - x_j \right)
= \pi \sum_{m=j}^{i-1} \zeta_{m}, (i > j)}
The permutations of the roots $x_i$'s act on $\zeta_i$'s as
the Weyl group of the type $A_{k}$.}

\smallskip\litem{{\it $D_{k}$ case.}}{In this case the fiber over the point $x$
is the rational curve $C_{x}$: $y^2 x + t_{0} y + R_{k-1}(x) + z^2 = 0$. 
Consider  the discriminant $\Delta (x) = t_{0}^2 - 4 x R_{k-1}(x), \,
R_{k-1} (x) = x^{k-1} + Q_{k-2}(x)$. Let $x_i$ be its roots: $\Delta (x_i)
= 0$.
For $x \neq x_i$ the rational curve  $C_{x}$ is isomorphic to $\IC^{*}$.}

{\leftskip=20pt
The period of the one-form $dy/2z$ is ${{\pi}\over{\sqrt{-x}}}$ for such an
$x$. 
The two-form
is given by:
$\omega_{\zeta} = {{dx \wedge dy}\over{2z}}$, hence its periods are:\par}
\eqn\prdsd{\left[ \omega_{\zeta} \right]_{ij} = i \pi \left( \sqrt{x_i} - 
\sqrt{x_j} \right)
= \pi \sum_{m=j}^{i-1} \zeta_{m}, (i > j)}
{\leftskip=20pt
The branching of the square roots in \prdsd\ and the permutations of $x_i$'s
generate  the action of the Weyl
group of the type $D_{k}$.\par}

\smallskip
\litem{{$E_{k}$ \it cases.}}{In these cases the fiber over $x$ is the elliptic
curve $y^3 + A(x) y + B(x) + z^2 = 0$ and it degenerates over the roots
$x_i$
of the discriminant $\Delta (x) = 4 A^{3}(x) + 27 B^2(x)$.
The periods are given by
$$
\int_{x_{i}}^{x_{j}} \gamma (x) dx
$$
where $\gamma (x) = \oint_{C_{x}} {{dy}\over{2z}}$, $C_{x}$ is the
one-cycle
which vanishes both at $x_i$ and $x_j$. So in this case the identification
between
the coordinates $t_{k}$ and $\zeta_{l}$ requires inverting the elliptic
functions.}

\subsec{Single $D$-brane at the generalized conifold
singularity}
We now proceed with describing $\CN=1$ theories whose Higgs branch coincides
with the non-compact Calabi-Yau manifold
 $Y_{\Gamma}$ with the conifold-like singularity
of the following type:
\eqn\cyt{F_{\Gamma} (\phi, x,y,z) \equiv \phi^{h} f_{\Gamma}\left( 
{{x}\over{{\phi}^{\alpha}}},
{{y}\over{{\phi}^{\beta}}}; t\right) + z^{2} = 0}
The equation \cyt\ is homogeneous with respect to the
$\IC^{*}$ action described in  \wght\ iff the variable $\phi$
has weight $1$.

First of all we need to show that these manifolds have shrunken three-cycles. 
Let us deform
the equation \cyt\ to
$$
\mu = \phi^{h} f_{\Gamma}\left( {{x}\over{{\phi}^{\alpha}}},
{{y}\over{{\phi}^{\beta}}}; t\right) + z^{2}
$$
Let us call the non-compact manifold described by this equation as
$Y_{\Gamma}(\mu)$. 
By construction the manifold $Y_{\Gamma}(\mu)$ is fibered over $\phi$ plane
with fiber over given $\phi$ being a particular ALE space
$S_{\zeta (\phi, \mu)}$. It is endowed with a holomorphic three-form:
\eqn\thrfr{
\Omega = {{d\phi \wedge dx \wedge dy \wedge
dz}\over{dF_{\Gamma}}}}
We are going to show that its periods scale as $\mu^{2\over h}$ and
therefore
vanish in the limit $\mu \to 0$. Indeed, the function $F_{\Gamma}$ is
homogeneous
of degree $h$ with respect to the $\IC^*$ action in \wght. Therefore the
form
$\Omega$ scales as $t^{2\over h}$ under the action of the element $t \in
\IC^*$.
Now let us turn to concrete examples of $A,D$ series.

The space $S_{\zeta(\phi, \mu)}$ is fibered over $x$-plane with
the generic 
fibers being isomorphic to either $\IC^{*}$ in the $A, D$ cases or elliptic
curves (with infinity deleted) in the $E$ cases. For given $\phi, \mu$ let
us
fix a one-dimensional contour connecting $x_i$ and $x_j$ over which the 
one-cycles vanish. As we vary $\phi$ these one-dimensional contours span a 
two-dimensional surface. The nontrivial three-cycle is obtained if we get
an interval in $\phi$ plane which connects two points $\phi_{ij}^{\pm}$
over which the zero-cycle $\left[ x_i \right] - \left[ x_j \right]$ shrinks
to zero (i.e. the points collide).  In the $A_k$ and $D_k$ cases we
can be explicit:

\litem{{\it  $A_{k}$ case.}}{The conditions on $\phi_{ij}^{\pm}$ are:
$$
P_{k+1} \left( {x\over {\phi}} \right) = {{\mu}\over{\phi^{k+1}}},
\quad 
P^{\prime}_{k+1} \left( {x\over {\phi}} \right) = 0 
$$
Let $\xi = x/{\phi}$. Then the period of the three-form
$\Omega$
reduces to
$$
- 2 \mu^{2\over{k+1}} \oint_{\sigma} \xi {{dw}\over{w^3}}
$$ 
where $\sigma$ is a non-trivial one-cycle on the curve 
$$
w^{k+1} = P_{k+1}(\xi; t)
$$ 
As $\mu \to 0$ all these periods clearly go to zero. 
}

\smallskip
\litem{{\it $D_{k}$ case.}}{Let $\xi = x/{\phi}^2, \eta = y/\phi$. 
Consider the curve:
$$
w^{k-1} = R_{k-1}(\xi; t) - {1\over{4\xi}} t_{0}^2
$$ 
The periods of the three-form $\Omega$ reduce to:
$$
- \mu^{1\over{k-1}} \oint \sqrt{\xi} {{dw}\over{w^2}}
$$
$\omega_{\zeta} = {{dx \wedge dy}\over{2z}},$ 
and they also vanish in the limit $\mu \to 0$. 
}

\smallskip
Now we wish to show that the manifold $Y_{\Gamma}$ is nothing but
the Higgs branch of the $\CN=2$ theory described above
perturbed by the superpotential term:
\eqn\pertu{W \to W - \sum_{i} {\half} m_{i} {\Tr} \phi_{i}^{2} }
with the only condition $\sum_{i} m_{i} = 0$.

Indeed, let us look at the equations $dW = 0$. By varying with respect
to the matter fields we get the condition
that $\phi_{i}$ must generate a trivial gauge transformation which is only
possible when:
\eqn\slp{\phi_{i} = \phi \Id_{n_{i}}}
Then, varying with respect to $\phi_{i}$ we get:
\eqn\mmp{\mu_{i} = - m_{i} \phi \Id_{n_{i}}}
The necessary and sufficient condition for the equations \mmp\
to be solvable is precisely $\sum_{i} n_i m_{i} = 0$ (it follows from
\TraceMu).\comment{depends a little on normalization of the $\mu_i$;
compare to \sprptn\ with factors of $n_i$.}
The space of solutions to \mmp\ is fibered over the $\phi \neq 0$
plane with the fiber being the (generically) smooth
ALE space,  corresponding to $\zeta_{i} = m_{i} \phi$. Thus the role
of the mass vector is to choose the direction in the moduli space of
ALE spaces of given ADE type.

In other words, the spaces $Y_{\Gamma}$ are constructed as follows.
Let $f(x,y)$ be any isolated simple singularity. Let $T$ be the
base of its versal deformation. The dimension $r$ of $T$ is called
the Milnor index of the singularity. It also coincides
with the  rank of the corresponding ADE group. The space $T$ has a natural
action of the $\IC^{*}$ group, which originates in the $\IC^{*}$ action
described in 
\wght. The space of orbits of this action is a weighted projective
space
$W\IP_{\{ {{d_{i}}\over{h}}\} }$  where
$d_{i}$ are the exponents of $f$ \arnold.
The space $T$ comes with the canonical bundle $Y$  (called Milnor
bundle). Its fiber $Y_{t}$ over point $t \in T$ is the surface $f(x,y; t) = 0$.

Choose any orbit $t = t(\phi)$ of the $\IC^{*}$ action. Restrict $Y$ onto
this orbit. This is our space $Y_{\Gamma}$. It depends on the choice of
orbit, that is on the choice of mass parameters $m_{i}$. 

\subsec{Relation to the geometric invariant quotient}

In this section we wish to show (in the $A$ and $D$ cases explicitly)
that the Higgs branch of the $\CN=2$ theory in the case where all
deformation
parameters are zero is nothing but the orbifold $\IC^2 /{\Gamma}$.
To this end we slightly reformulate the solution for the $F$-flatness
conditions.
Let $V = \IC^{\Gamma}$ - the space of $\IC$-valued functions on the group
$\Gamma$.
This is naturally a representation (called regular representation)
 of $\Gamma$, induced for concreteness by
the
left action of $\Gamma$ on itself. For $g \in \Gamma$ let $\gamma (g)$ be
the
corresponding element of ${\rm Hom}(V,V)$. Consider the space of pairs
$X_{\alpha} \in {\rm Hom}(V, V)$,
$\alpha = 1,2$ of operators in $V$ which obey two conditions:
\eqn\cond{\eqalign{[X_{1}, X_{2}] & = 0 \cr
g_\alpha{}^\beta X_{\beta} & = \gamma (g) X_{\alpha} \gamma(g)^{-1}\cr}}
where $g_\alpha{}^\beta$ are the matrix elements of $g$ in the
two-dimensional
representation of $\Gamma$. Our space is the quotient of the space of these
pairs $(X_1, X_2)$ by the action of the group of gauge transformations
(cf. \dm).
The latter are the elements of ${\rm End}(V)$ which commute 
with $\gamma(g)$ for all $g \in \Gamma$. 

Now let $f(z_1, z_2)$ be any $\Gamma$-invariant
function on $\IC^2$. Consider the matrix $f(X_1, X_2)$.
Due to invariance of $f$ we have:
\eqn\inva{
\gamma (g) f(X_1, X_2) \gamma (g)^{-1} = f(X_1, X_2) }
for any $g \in \Gamma$.

Now let us look at the $A_{k}, D_{k}$ examples in some detail.

\litem{{\it $A_{k}$ case.}}{In this case the solution for $(X_1, X_2)$ is:
\eqn\explc{X_{1} = z_{1} J_{+}, \quad X_{2} = z_{2} J_{-}}
where
\eqn\shif{J_{+} = \pmatrix{0 & 1 & 0 & \ldots & 0\cr
0 & 0 & 1 & \ldots & 0 \cr
\vdots & \vdots & \vdots &  & 1\cr
1 & 0 & 0 & \ldots & 0\cr}, \quad J_{-} = J_{+}^{t}}
in the basis where $\gamma (g) $ is diagonal matrix
 with entries
being $1, \omega, \omega^2, \ldots, \omega^{k}$, $\omega = e^{{2\pi
i}\over{k+1}}$.
The basic invariants are: $\CX = z_1^{k+1}, \CY = z_2^{k+1}, \CZ = z_1
z_2$, which obey the equation with $A_{k}$ singularity:
$$
\CX \CY = \CZ^{k+1}
$$
The corresponding matrix functions are clearly scalars: 
$$
X_{1}^{k+1} = \CX \cdot \Id , X_{2}^{k+1} = \CY \cdot \Id, X_1 X_2 = \CZ
\cdot \Id
$$ 
}

\smallskip
\litem{{\it $D_{k}$ case.}}{In this case the matrices $X_{1}, X_{2}$
have a block diagonal form:
\eqn\explcd{X_{1} = \pmatrix{z_{1} J_{+} & 0 \cr 0 & i z_{2} J_{+}\cr},
\quad 
X_{2} = \pmatrix{z_{2}J_{-} & 0 \cr 0 & i z_{1} J_{-} \cr }}
where the size of $J_{\pm}$ is $2(k-2) \times 2(k-2)$.
The basic invariants here are:
\eqn\bscinv{\CX = z_{1}^{2(k-2)} + (-)^{k} z_{2}^{2(k-2)}, \,  \CY = z_1 z_2
\left( z_{1}^{2(k-2)} - (-)^{k} z_{2}^{2(k-2)} \right), \, \CZ = z_1^{2}
z_2^{2} }
which obey $D_{k}$-type equation:
$$
\CY^2 = \CZ \CX^2 - 4 (-)^{k} \CZ^{k-1}
$$
It is obvious that 
$$
\CX (X_1, X_2) = \CX \cdot \Id, \, \CY (X_1, X_2) = \CY \cdot \Id, \,  \CZ
(X_1, X_2) = \CZ \cdot \Id
$$
}
\smallskip\noindent
The matrices $X_1,X_2$ provide the most efficient way of making
completely explicit the abstract construction of the spaces $Y_\Gamma$
which we sketched at the end of section~\FirstConstruct.  If we
introduce a third matrix $\Phi$ with $\gamma(g) \Phi \gamma(g)^{-1} = \Phi$,
then the superpotential is
  \eqn\MatrixSuperPotential{
   W = \Tr \left( \Phi [X_1,X_2] - \tf{1}{2} M \Phi^2 \right) \ .
  }
 The requirement of F-flatness is
  \eqn\FFlatMatrix{\eqalign{
   &[\Phi,X_1] = 0 \qquad [\Phi,X_2] = 0  \cr
   &[X_1,X_2] = M \Phi \ .
  }}
 The first two expressions in \FFlatMatrix\ are satisfied when $\Phi$
is a trivial gauge transformation: $\Phi = \phi \Id_{|\Gamma|}$ for some
complex number $\phi$.  Taking the trace of the last equation in
\FFlatMatrix\ tells us $\Tr M = 0$.  The space of
solutions to this equation
modded out by the complexified gauge group (which implements
D-flatness along with gauge invariance) should be the generalized
conifold.

\litem{{\it $A_k$ case.}}{Let us use the notation
  \eqn\DiagNote{
   \diag{x_i} = \diag{x_i}_{i=1}^\Gamma = \diag{x_1,x_2,\ldots,x_\Gamma}
    = \pmatrix{x_1 & 0 & \ldots & 0 \cr
               0 & x_2 & \ldots & 0 \cr
               \vdots & \vdots & & \vdots \cr
               0 & 0 & \ldots & x_\Gamma}
  }
 for diagonal matrices.  So for instance $M = \diag{m_i}$.  As a
subgroup of $SU(2)$, $A_k$ has as its generating element
  \eqn\GenerateAk{
   (g_1)_\alpha{}^\beta = 
    \pmatrix{\omega & 0 \cr 0 & \omega^{-1}}_\alpha^\beta \ .
  }
 In a basis for the regular representation where $\gamma(g_1) =
\diag{1,\omega,\ldots,\omega^k}$, the general solution to
$(g_1)_\alpha{}^\beta X_\beta = \gamma(g_1) X_\alpha \gamma(g_1^{-1})$
is $X_1 = \diag{b_{i,i+1}} J_+$, $X_1 = J_- \diag{b_{i+1,i}}$.  We
have $X_1 X_2 = \diag{b_{i,i+1}b_{i+1,i}}$ and $X_2 X_1 =
\diag{b_{i-1,i} b_{i,i-1}}$.  To satisfy $[X_1,X_2] = M\Phi$ we must
have $X_1 X_2 = x \Id_\Gamma - \Xi \phi$ for some complex number $x$
and a matrix $\Xi = \diag{\xi_i}$ where $\xi_{i-1}-\xi_i = m_i$.  By
convention we may take $\xi_i = -\sum_{j=1}^i m_i$.  Defining the
gauge invariant quantities $c_+ = \det X_1$, $c_- = \det X_2$, we
recover the $A_k$ conifold equation from
  \eqn\AkConifold{
   c_+ c_- = (\det X_1)(\det X_2) = \det (X_1 X_2) = 
    \prod_{i=1}^{k+1} (x - \xi_i \phi) \ .
  }
 It is easy to understand this point how the F-flatness conditions
eliminate what seems {\it a priori} to be an excess of gauge-invariant
products parameterizing the moduli space.  In addition to the products
$c_\pm = \prod_i b_{i,i\pm1}$ which take us all the way around the
extended Dynkin diagram, there are $k+1$ products
$b_{i,i+1}b_{i+1,i}$.  But these may all be expressed in terms of $x$
and $\phi$, with \AkConifold\ being the only equation among $x$,
$\phi$, and $c_\pm$, so indeed the moduli space has three complex
dimensions.
}

\subsec{Construction of the Kahler metric on $Y_{\Gamma}$}

Gauge theory gives us an explicit construction of
the Kahler metric on the space $Y_{\Gamma}$. Of course, in the
case $m=0$ the metric is exact, while in the $m \neq 0$ case it may
be affected by the quantum corrections which lead
to the renormalization of the Kahler potential. At any rate we shall describe 
the
metric which one gets by the Kahler quotient construction in the $A_{k}$ case.

The original space of fields is $\CB = \{ ( \phi_{i}, b_{i, i+1}; b_{i+1, i} )
\vert\ i = 0, \ldots, r\}$, where all fields are complex and we have identified 
$r+1 \equiv 0$.  We assume that the metric on $\CB$ is flat:
\eqn\flt{
ds^{2} = \sum_{i=0}^{k} d\phi_{i} d\bar\phi_{i} + db_{i, i+1}d\bar b_{i, i+1}
+ db_{i+1, i} d\bar b_{i+1, i \ .}
}
We now impose the $D$- and $F$-flatness conditions, which means that we
restrict the metric \flt\ onto the surface of equations:
\eqn\equu{\eqalign{& \vert b_{i,i+1} \vert^{2} - \vert b_{i+1, i} \vert^{2}
+ \vert b_{i, i-1} \vert^{2} - \vert b_{i-1, i} \vert^{2} = 0 \cr
&b_{i,i+1}b_{i+1, i} - b_{i, i-1}b_{i-1, i} = m_{i} \phi_{i}\cr
& b_{i, i+1} ( \phi_{i} - \phi_{i+1} ) = 0 \cr
& b_{i+1, i} ( \phi_{i+1} - \phi_{i} ) = 0\cr}}
It is convenient, following \girych\ to rewrite the flat metric
on $b$'s in terms of the coordinates $(\vec r_{i}, \theta_{i})$, where
\eqn\hypk{\eqalign{& \vec r_{i} = (t_{i}, x_{i}, \bar x_{i}) \cr
& \vert b_{i,i+1} \vert^{2} - \vert b_{i+1, i} \vert^{2}  = 2t_{i}\cr
& b_{i, i+1}b_{i+1, i} = x_{i}\cr
& b_{i, i+1}/ b_{i+1, i}  =  \vert b_{i, i+1}/ b_{i+1, i} \vert 
e^{2i\theta}\cr}}
We have:
\eqn\flti{ db_{i+1, i} d\bar b_{i+1, i} + db_{i, i+1}d\bar b_{i, i+1} = 
{1\over{r_{i}}} d\vec r_{i}^{2} + r_{i} \left( d\theta_{i} + \omega_{i}
 \right)^{2}}
where $r_{i} = \vert \vec r_{i} \vert$, and the Dirac connection $\omega_{i}$ 
obeys:
$$
d\omega_{i} = \star d {1\over{r_{i}}}
$$
where $d$ is three-dimensional and $\star$ is taken with respect to
the flat metric on $\IR^{3}$. The gauge group acts as follows:
\eqn\ggr{\theta_{i} \mapsto \theta_{i} + \alpha_{i} - \alpha_{i+1}}
The $D,F$-flatness conditions imply that:
\eqn\slv{\eqalign{& \phi_{0} = \phi_{1} = \ldots = \phi_{k} = : \phi\cr
& t_{0} = t_{1} = \ldots = t_{k} = : t\cr
& x_{i}  = x - \xi_{i} \phi\cr}}
where $\xi_{i}$ are the complex numbers which are uniquely
specified by the following conditions:
\eqn\cndx{\xi_{i} - \xi_{i-1} = -m_{i},\qquad \sum_{i} \xi_{i} = 0}
The projection  along the orbits of the gauge group is achieved
by taking the orthogonals to the orbit. Formally this is equivalent to
the following procedure \girych\ : replace $d\theta_{i}$ by $d\theta + A_{i} - 
A_{i+1}$,
compute the $ds^{2}$ and minimize with respect to $A_{i}$. The result is
the following metric:
\eqn\fnmtrc{ds^{2} = V d\vec r^{2} - U d\phi d \bar x  - 
\bar U  d \bar\phi d x +  W d \phi d\bar \phi + 
V^{-1} 
\left( d\theta + A \right)^{2}  }
where
\eqn\fnmtri{\eqalign{& V = \sum_{i=0}^{k} {1\over{ \sqrt{ t^{2} + \vert x  - 
\xi_{i} \phi \vert^{2}}}}\cr
& U = \sum_{i=0}^{k} {{{\xi}_{i}}\over{ \sqrt{ t^{2} + \vert x - \xi_{i} \phi 
\vert^{2}}}} \cr
& W = \sum_{i=0}^{k} \left( 1 + 
 {{\vert  {\xi}_{i} \vert^{2} }\over{ \sqrt{ t^{2} + 
\vert x  - \xi_{i} \phi \vert^{2}}}}\right)  \cr
& d A = \star d V\cr}}
where in the last formula $d$ is taken with respect to $(t, x, \bar x)$.

\subsec{Properties of the
metric on $Y_{\Gamma}$}

The space $Y_{\Gamma}$ comes equipped with the holomorphic three-form.
In the $A_{k}$ case it is given by the formula:
\eqn\hlm{\Omega = {{dx \wedge d\phi \wedge d c^{+}}\over{c^{+}}}}
where $c^{\pm} = y \pm i z$. In terms of the coordinates $b_{i,i+1}$ etc.
the variables $c^{\pm}$ are expressed as follows:
\eqn\expr{c^{\pm} = \prod_{i=0}^{k} b_{i, i\pm 1}}
In solving the $D,F$-flatness conditions a choice of the
gauge for the phases of $b_{i,i\pm1}$ has to be made.
We choose:
\eqn\ggge{\eqalign{
& b_{i,i+1} = \left[ t + \sqrt{t^{2} + \vert x - \xi_{i} \phi \vert^{2}} 
\right]^{1\over 2} e^{{i\theta}\over{k+1}} \cr
& b_{i+1, i} = \left[ - t +  \sqrt{t^{2} + \vert x - \xi_{i} \phi \vert^{2}} 
\right]^{1\over 2} {{x - {\xi}_{i} \phi}\over{\vert x - \xi_{i} \phi \vert}}
e^{-{{i\theta}\over{k+1}}}\cr}}
With this choice of phases the Kahler form $\varpi$ on $Y_{\Gamma}$
is written out as follows:
\eqn\klhr{\varpi = dt \wedge \left( d\theta + A\right) +  {i\over{2}}
d\phi \wedge d\bar\phi + 
{{i}\over{4}} \sum_{l=0}^{k} {{d ( x - \xi_{l} \phi) \wedge d {\overline{( x - 
\xi_{l} \phi)}}}\over{\sqrt{t^{2} + \vert x - \xi_{l} \phi \vert^{2}}}}}
where 
\eqn\frlma{A = {\half} \sum_{l}  \left( {{t}\over{\sqrt{t^{2} + \vert x -  
\xi_{l}\phi\vert^{2}}}} - 1 \right) d \left( {\rm arg} ( x - \xi_{l}\phi 
)\right)  }
Of course, another choice of gauge leads to the gauge transformed $A$.
Now, the holomorphic three-form turns out to have rather simple form:
\eqn\hlmr{\left( i ( d\theta +A) + {\half} V dt \right) \wedge dx \wedge d\phi}
{}From this expression we get the volume form:
\eqn\vlm{\Omega \wedge \bar \Omega = -i V d\theta \wedge dt \wedge dx \wedge
d\bar x \wedge d\phi \wedge d\bar \phi }

For the Kahler metric to be Ricci-flat it is necessary that:
\eqn\rcf{
\varpi \wedge \varpi \wedge \varpi = {\kappa} \Omega \wedge \bar\Omega}
for some constant $\kappa$.
Explicit computation shows that :
\eqn\expl{
\varpi^{3} = {3\over{8}} \left( V W - \vert U \vert^{2} \right)
dx \wedge d\bar x \wedge d\phi\wedge  d\bar\phi \wedge d\theta \wedge dt }
and \rcf\ is not obeyed. Instead, for the Ricci tensor we have:
\eqn\rcci{R \equiv R_{i\bar j} dz^{i} \wedge dz^{\jb} = 
\p \pb {\rm log} \left( W - {{\vert U \vert^{2}}\over{V}} \right)}

The fact that the metric doesn't come out in the Ricci-flat form
may sound troubling. 
On the other hand it seems  that it does not receive quantum loop
corrections. 
The reason is that we study abelian gauge theory and the coupling in
this
theory is weak in the infrared so all loops must go away.

\lref\dg{M.~Douglas, B.~Greene, ``Metrics on D-brane orbifolds'', 
hep-th/9707214, Adv. Theor. Math. Phys. {\bf 1} (1998) 184-196.}

Hence we are led to believe that, in contrast to $\CN=2$ case, here the
geometry as observed by the single $D3$-brane and by the fundamental string
is different---unless the assumption \flt\ that the initial Kahler metric
was flat is wrong.  See \dg\ for a thorough discussion, and also \gdm\ for
more examples of $\CN=1$ orbifolds.

Another problem is that different terms in the formula for the metric \fnmtrc\
and the Kahler form \klhr\ scale differently under
the $\IR_{+}$ which is a part of the $R$-symmetry \wght.

\subsec{Geometry  of the base of the cone}
\subseclab\BaseGeom

Nevertheless, close to the singularity where $t= \vert x \vert = \vert
\phi \vert = 0$ the term $d\phi \wedge d\bar\phi$ can be neglected.
As a result of this `` RG flow '' the metric on the Higgs
branch
becomes invariant under ``RG'' action of
$\IR_{+}$. Moreover, the Higgs branch becomes a cone over a fivefold
$M^{5}$ which is in turn a $U(1)$ bundle over four dimensional
Kahler manifold $B$ which we now describe in some detail.

Let us think of $Y_{\Gamma}$ as the symplectic manifold with the
two-form $\varpi$. It is invariant under the $U(1)$ action
$\phi \mapsto \phi e^{i\alpha}, x \mapsto x e^{i\alpha}, \theta \mapsto
\theta + {{k+1}\over{2}} \alpha$. This action is generated by the Hamiltonian:
\eqn\ham{H = {1\over{2}} \vert \phi \vert ^2 + {1\over{2}} \sum_i 
\sqrt{t^2 + \vert x - \xi_i \phi \vert^2}}
It is easy to show using the metric \fnmtrc\ that this action is also
free. So we have a fiber bundle:
\eqn\bndl{\matrix{U(1) & \rightarrow & Y_{\Gamma} \cr \quad & \quad & \downarrow 
\cr
\quad & \quad & B \cr}}
where the manifold $B$ is described as a quotient of the subvariety in
$Y_{\Gamma}$
defined by the equation 
$H = \zeta  > 0$ by the action of the $U(1)$.
The base of the cone $M^5$ is the level set of the Hamiltonian:
$H = \zeta$. To be more precise, consider the following ``RG flow'':
perform the simultaneous rescaling of $t, \vert x \vert , \vert \phi
\vert$
and $\zeta$ by the same amount $\mu$ and then take the limit
$\mu \to 0$. As a result we get the manifold
$M^{5}$, defined as the hypersurface 
\eqn\hyper{\sum_i \sqrt{t^2 + \vert x - \xi_i \phi \vert^2} = 1} 
in the space of
$t, \theta, x, \phi$, with the metric
\fnmtrc\ where $W \to W - (k+1)$. In the case $k=1$ the manifold $B$
is the set of pairs of vectors $\vec x, \vec y, \, \vec y \in \IR^3,
\vec x \in \IR^2 \subset \IR^3$ subject to the condition
$\vert \vec x - \vec y \vert + \vert \vec x + \vec y \vert = 1$.
It is easy to show that this space is isomorphic to $S^2 \times S^2$
in agreement with the expectations about the conifold geometry
(cf.
\witkleb).

In general the manifold $B$ can be described as a complex hypersurface
in the
weighted projective space $W\IP_{1, \a, \b, {h \over 2}}^{3}$ defined by the
equation
$F_{\Gamma} = 0$. For example, in the $A_{1}$ case we would get
the hypersurface in the ordinary $\IP^{3}= \{ ( X : Y : Z :T ) \}$ 
defined by the equation
$X^2 + Y^2 + Z^2 + T^2 = 0$ that is the quadric $\IP^1 \times \IP^1$.

\newsec{Supergravity and holography}
\seclab\Supergravity

\subsec{Large number $N$ of $D$-branes at the singularity: Gauge theory}
\subseclab\GenAllN

It is clear how to proceed with generalizations:
replace the vector spaces $\r_{i}$ by $R_{i} = \IC^{N} \otimes
\r_{i}$. The gauge group is now:
\eqn\ggrii{G_{N} = \times_{i} U(N n_{i})}
The matter fields are the $a_{ij}$
hypermultiplets
in the  bi-fundamental representations:  $(Nn_{i}, {\overline{Nn_{j}}})$.

This $\CN=2$ theory describes $N$ coincident $D3$-branes placed
at the orbifold point in $\IC^{2}/\Gamma$. 
We now perturb this theory by adding the mass term:
\eqn\prtb{W \to W - \sum_{i} \half m_{i} {\Tr} \phi_{i}^{2}} 
In the infrared limit the $U(1)$ factors decouple and one is left with the
gauge group
\eqn\ggrp{{\tilde G}_{N} = \times_{i} SU(Nn_{i})}
and the effective superpotential:
\eqn\efsp{W_{eff} = \sum_{i:\, m_i \neq 0} {1\over{2m_{i}}} {\Tr} \mu_{i}^{2}
+ 
\sum_{i:\, m_i = 0} {\Tr} \mu_i \phi_i}

Methods described in \LeighStrassler\ can be used to determine the possible
anomalous dimensions at the infrared fixed point.  Linear constraints on
the anomalous dimensions of chiral fields result from setting the NSVZ
exact beta functions to zero and demanding that the superpotential be
dimension~$3$.  These constraints can always be satisfied in our models by
giving the $\phi_i$ anomalous dimensions of $1/2$ and the $B_{ij}$
anomalous dimensions of $-1/4$.  Typically the number of independent
constraints is less than the number of independent anomalous dimensions, so
there is actually a space of solutions.  When all $m_i$ are non-zero, it is
straightforward to see from the condition on the superpotential that the
dimensions of all gauge-invariant combinations of the $B_{ij}$ are
invariant over this space.  Thus we can calculate these dimensions at the
point where all the $B_{ij}$ have anomalous dimension $-1/4$.  This is the
result we actually will use in comparisons with supergravity predictions.
By continuity we would expect it to continue to hold as some of the $m_i$
are taken to zero (but not all, since the $m_i$ are only defined up to an
overall rescaling).  We do not have a proof of this, but we would be
surprised to find a continuous spectrum of possible dimensions for gauge
invariant operators.


We now proceed with showing that UV superpotential yields the
moduli space of $N$ $D3$-branes placed at $Y_{\Gamma}$. 

Indeed, the discussion of the section related to the geometric invariant
theory goes through with the only change that we now tensor $\IC^{\Gamma}$
by the
dummy space $\IC^{N}$ which is a trivial representation of the group
$\Gamma$.
As a consequence, the expressions for
$X_{1,2}$ now have the following form:
$$
X_{\alpha}^{(N)} = {\rm diag}\left( X_{\alpha}^{(1)} ( z_{1}^{1},
z_{2}^{1}),
\ldots, X_{\alpha}^{(1)}(z_{1}^{N}, z_{2}^{N}) \right)
$$
so they depend on an $N$-tuple of the parameters $z_{1}, z_{2}$
on which 
the allowed gauge transformations act as the group $\Gamma$ does. 
Turning on the mass matrix $M$ makes $z_1, z_2$ live in the
deformed ALE space. It means
that
the Higgs branch looks like the $N$'th symmetric product of the
generalized conifold $Y_{\Gamma}$, which is what we expect.

\subsec{The supergravity geometry and chiral primaries}
\subseclab\GeomCP

Let us start by briefly reminding the reader the supergravity version
of D3-branes at an isolated singularity of a six dimensional manifold
\kehag.  Assume the manifold is a Calabi-Yau three-fold and that the
metric near the singularity before the addition of D3-branes can be
written as
  \eqn\ConeGeom{
   ds_6^2 = dr^2 + r^2 g_{\alpha\beta} dx^\alpha dx^\beta
  }
 where $g_{\alpha\beta}$ is an Einstein metric on a five-manifold
$M^5$.  When $N$ $D3$-branes are placed at the singularity $r=0$, the
supergravity metric is
  \eqn\KGeom{
   ds^2 = \left( 1 + {L^4 \over r^4} \right)^{-1/2} 
    (-dt^2 + dx_1^2 + dx_2^2 + dx_3^2) + 
    \left( 1 + {L^4 \over r^4} \right)^{1/2} ds_C^2
  }
 where 
  \eqn\LReq{
   L^4 = {\sqrt\pi \over 2} {\kappa N \over \Vol M_5} \ .
  }
 Here $2\kappa^2 = 16 \pi G = (2\pi)^7 g_s^2 \alpha'^4$ is the
gravitational coupling.  Supergravity is a good approximation when $L$
is much bigger than the the Planck length and the string length.  

This pictures applies to the conifolds \cyt\ as follows.  In \wght\ we
specified an action of $\IC^*$ action on the conifold geometries.  The
$\IR^+$ part of this action is the dilation symmetry of the cone, $r
\to \lambda r$.  The $U(1)$ part of the action is an isometry of
$M_5$, and in the field theory it is realized as the $R$-symmetry
group.  From the existence of the Calabi-Yau metric on the conifolds
we are learning of the existence of a class of five-dimensional
Einstein manifolds.  Unlike the coset manifolds constructed in
\romans, these Einstein manifolds have moduli spaces.  To discuss
chiral primary operators in following \gEin\ is impractical because we
cannot write down the metric explicitly.  Fortunately there is a more
efficient way, which we will now explain.

A complete set of harmonic functions on the cone can be generated from
harmonic functions $f$ which are also eigenfunctions of the operator
$r\partial_r$:
  \eqn\Laplap{\eqalign{
   r\partial_r f &= \Delta_f f  \cr
   2 (\bar\partial\bar\partial^* + \bar\partial^*\bar\partial) f &= 
    (d d^* + d^* d) f =
    \left[ {1 \over r^5} \partial_r r^5 \partial_r + 
     {1 \over r^2} \square \right] f = 0 \ ,
  }}
 where we have reserved the symbol $\square$ to denote the
five-dimensional Laplacian on the base of the cone.  Together the two
conditions in \Laplap\ imply that $(\square+E) f = 0$ where $E
= \Delta(\Delta+4)$.  By considering a complete set of harmonic
functions on the cone one can extract the full spectrum of the
five-dimensional scalar Laplacian.

The holographic correspondence as worked out in \gkp\witadsi\ relates
on-shell fields in the bulk of spacetime to operators on the boundary.
In the present context, following the arguments used in \gEin, the
spectrum of the scalar Laplacian $\square$ relates to chiral primary
operators with dimension $\Delta = -2 + \sqrt{4+E}$: exactly the
eigenvalue under $r\partial_r$ of the harmonic extension $f$ to the
cone!  To be more precise, only a subset of the eigenfunctions of
$\square$ correspond to chiral primaries: these are the operators
which maximize $U(1)_R$ charge with dimension held fixed, and the
eigenfunctions they are dual to are in fact the ones which extend to
holomorphic functions on the cone.

In particular, for the $A_k$ conifolds, we can consider the complex
variables $c^\pm$, $x$, and $\Phi$ as holomorphic functions.  Near the
IR fixed point we know their dimensions from their representations as
products of the fields $b_{i,i\pm1}$: $\Delta_{c^\pm} = {3\over
4}(k+1)$ and $\Delta_x = \Delta_\Phi = 3/2$.  These dimensions are
just $3/2$ times the $R$-charges listed in \wght.  Since these
$R$-charges are determined by the $\IC^*$ action on the conifold
geometry, of which $r\partial_r$ is one generator, we have shown that
the dimensions of $c^\pm$, $x$, and $\Phi$ agree between gauge theory
and supergravity, up to an overall normalization.  The most direct way
to fix that normalization is to note that the metric \ConeGeom\ has
dimension $2$ under dilations of $r$; hence so does the Kahler
form.\foot{We note in passing that an appropriately chosen Kahler
potential should also have dimension $2$.  Written in terms of the
complex variables, this amounts to the homogeneity condition
  \eqn\KScale{
   \sum_i \Delta_i \left( {\partial \over \partial\log z_i} + 
    {\partial \over \partial\log \bar{z}_i} \right) K = 2K \ ,
  }
 where we collectively denote $c^\pm$, $x$, and $\Phi$ by $z_i$.}  The
cube of the Kahler form is proportional to $\Omega \wedge \bar\Omega$,
where $\Omega$ is the complex three-form.  So $\Omega$ has dimension
$3$.  Finally, writing out $\Omega$ in terms of $c^\pm$, $x$, and
$\Phi$, one can verify that the gauge theory dimensions indeed agree
completely with supergravity.  Although we have focused on the $A_k$
conifolds the analysis is equally straightforward for the $D_k$ and
$E_k$ cases.

Holomorphic functions of the $z_i$ which have a definite eigenvalue
under $r \partial_r$ are just polynomials in the $z_i$ which are
homogeneous with respect to the weight in \wght, identified modulo the
equation relating the $z_i$ which defines the conifold.  We have
argued in section~\GaugeTheory\ that the solution of the F-flatness
conditions in the gauge theory, modulo complexified gauge invariance,
results in this same conifold.  The complexification of the gauge
invariance implemented D-flatness.  Now, chiral primaries in the gauge
theory are constructed from sums of gauge invariant products of chiral
superfields, modulo F- and D-flatness conditions, and with definite
conformal dimension.  It follows that chiral primary operators are in
one-to-one correspondence with the homogeneous holomorphic functions on
the conifold.  This was essentially checked in (34) of \witkleb\ for
the $A_1$ case by showing explicitly that F-flatness constraints
allowed one to symmetrize $A_i$ and $B_j$ fields separately in a
product of the form $A_{i_1}B_{j_1}A_{i_2}B_{j_2}\ldots
A_{i_l}B_{j_l}$.  The constraints are more complicated in the general
$ADE$ case, but the conclusions are the same: because the moduli space
(namely the conifold) is parametrized by holomorphic gauge invariant
combinations of the matter fields modulo the F-flatness conditions,
the chiral primary fields which these combinations represent are
precisely the holomorphic functions on the conifold.

It is worth emphasizing that holomorphic functions on the conifold
were introduced as a trick to write down efficiently the
eigenfunctions of the Laplacian on $M_5$ which minimized dimension for
specified $R$-charge.  The arguments of the previous two paragraphs
show with minimal calculation that holography predicts exactly the
right dimensions and degeneracies for chiral primary operators in the
gauge theory.  There are on order $\Delta^3$ chiral primaries with
dimension less than $\Delta$.  As in the $A_1$ case \gEin\ (but in
contrast to the $S^5$ case) supergravity predicts in addition on order
of $\Delta^5$ non-chiral fields of dimension less than $\Delta$.
These come from the non-holomorphic eigenfunctions of the Laplacian on
$M_5$.  In the gauge theory they reside in long multiplets whose
dimensions are not algebraically protected, and as far as we know
there is no good understanding for why the dimensions should match the
supergravity predictions.

\subsec{Blowup Modes and RG flow}
\subseclab\BlowRG

If it is indeed true that one can {\it define} string theory on a
manifold which is (at least asymptotically) $AdS_5$ times a compact
manifold through a gauge field theory which lives on the boundary,
then we would expect to see reflected in some solution of supergravity
the full renormalization group flow from an $\CN = 2$ theory, deformed
by mass terms as in \pertu, to a non-trivial infrared $\CN = 1$ fixed
point with a quartic superpotential.  The simplest case would be a
supergravity geometry interpolating between $S^5/\IZ_2$ and $T^{11}$.
We do not have a complete enough understanding of the Lagrangian of
gauged $\CN = 4$ supergravity in five dimensions to find such
solutions explicitly.\foot{For a globally supersymmetric conformal
field theory in four dimensions, $\CN = 2$ means sixteen real
supercharges, which is the same number as in $\CN = 4$ gauged
supergravity in five dimensions.  The supergroup organizing the
multiplets in both cases is $SU(2,2|2)$.}  However, we can at least
describe a multiplet which plays a key role.

Blowup modes of the fixed $S^1$ of $S^5/\Gamma$ were discussed in
\gukov\ (see also the appendix of \GukKap\ for a more precise
discussion of Kaluza-Klein reduction).  Our aim is to indicate how
this analysis feeds into the supergravity interpretation of the RG
flow.  For $\Gamma = A_k$, $D_k$, or $E_k$, blowing up the $S^1$
introduces $k$ independent 2-cycles.  The self-dual four-form
potential $A^+_{MNPQ}$ on one of these cycles gives rise to an
anti-self-dual two-form potential $B^-_{MN}$ in $AdS_5 \times S^1$.
The Kaluza-Klein reduction of $B^-_{MN}$ on $S^1$ leads a tower of
fields labelled by the Kaluza-Klein momentum $\ell$: at $\ell=0$ a
vector field $A_\mu$ satisfying $d * d A = 0$; and for $\ell\neq 0$ an
antisymmetric tensor field $A_{\mu\nu}$ satisfying $*dA = -i\ell A$.
Both these equations of motion are valid only at the linearized level.
Tensor fields which satisfy the latter equation of motion are termed
``anti-self-dual'' in \grw, where (among other things) the
superpartners of anti-self-dual antisymmetric tensors and of vectors
are worked out using $SU(2,2|2)$ group theory.  The multiplet we will
be particularly interested in is the $\ell = 1$ tensor multiplet.  The
bosonic components, their quantum numbers under the R-symmetry group
$SU(2) \times U(1)$, and the types of gauge theory operators they are
dual to, are as follows.
  \eqn\TensorTable{
   \seqalign{\span\TT\quad & \span\TC\quad & \span\TC\quad & \span\TC}{
   field  & SU(2) \times U(1) & \hbox{operator} & \hbox{dimension}  \cr
   scalar & {\bf 1}_0 & F^2 & 4  \cr
   scalar & {\bf 1}_4 & X^2 & 2  \cr 
   scalar & {\bf 3}_2 & \lambda\lambda & 3 \cr
   tensor & {\bf 1}_2 & F_{\mu\nu} X & 3
  }}
 The gauge theory operators we have identified schematically as $X^2$
in \TensorTable\ can be written more precisely as $\tr \Phi_i^2 - \tr
\Phi_j^2$.  The set of scalar mass terms corresponding to all the
independent 2-cycles of the blown up orbifold form a basis for the
mass perturbations introduced in \pertu.  The $\lambda\lambda$
operators in \TensorTable\ are the corresponding fermion mass terms
which are turned on at the same time to preserve $\CN = 1$
supersymmetry.

 Part of the analysis of \gukov\ was to determine the
dimensions of these operators, listed in \TensorTable, from the masses
of the corresponding modes in the tensor multiplet.  The first step in
finding a supergravity solution interpolating between the orbifold and
conifold geometries should be to turn on the $\lambda\lambda$ and
$X^2$ modes in \TensorTable\ at the linearized level.  One would need
a concise description of the relevant interactions to extract a
solution of the full nonlinear theory, perhaps along the lines of \dz,
\ppz.  Unlike the RG flows considered in those papers, the
supergravity solution interpolating between the orbifold and the
conifold should preserve four real supercharges throughout the flow.
At the UV and IR endpoints one should recover sixteen and eight
supercharges, respectively.

The relevant part of the multiplet \TensorTable\ which is turned on
forms what is called a {\it spinor} multiplet of $\CN=2$ $AdS_5$
supergravity \grw. It contains a pair of scalars of $U(1)$ charges $\pm
1$ and a spinor (left- or right- handed) of $U(1)$ charge zero. The
$U(1)$ charge assignments can be shifted by the Kaluza-Klein momentum
$\ell$ of the highest spin state.  Thus in particular, two
right-handed spinor multiplets with $\ell = 0$ and $2$, together with
an anti-self-dual tensor multiplet of $\CN=2$ with $\ell = 1$, form
the anti-self-dual tensor multiplet of $\CN=4$ with $\ell=1$ that
enters into \TensorTable.

\newsec{Dual constructions with branes}
\seclab\DualBranes

Most of the constructions which we were studying using the geometry
or field theory can be redone in the language of branes, along the
lines of \uranga, \mukhi.
{}The idea is to use the $T$-duality between the ALE (more precisely
multi-Taub-NUT)
space and fivebranes. 
{}First let us remind the reader of the 
realization of the $\CN=2$ superconformal
theories
in this language. Consider $N$ $D3$-branes placed at the orbifold
singularity
of the ALE space. Let $0123$ be the world-volume of the $3$-branes,
while $6789$ are the coordinates of ALE space. Let $6$ be the compact
direction corresponding to the $U(1)$ isometry of the ALE space. 
Perform $T$-duality along the $6$'th direction. If the
ALE singularity is of $A_{k-1}$ type then we 
get the Type $\II$A theory on $\IR^{1,8} \times S^1$
with $k$ $NS5$-branes, whose
world-volume is $012345$ ($6$ being the coordinate along $S^1$) and which 
are located at the same point $\vec r$ in the $789$-plane. 
The $N$ $D3$ branes are mapped to the $N$ $D4$-branes which wrap the
circle $S^1$. Their world-volume is $01236$. The $NS5$-branes are
located at the points $\theta_1, \ldots, \theta_k$. The differences
$\theta_i - \theta_{i-1}$ correspond to the fluxes of the
$NSNS$ $B$-field on the Type $\II$B side. The corresponding
$RR$ fluxes become visible if the whole picture is lifted to
$M$-theory,
where the circle $S^1_{6}$ is promoted to the two-torus
$T^2 = S^1_{6} \times S^{1}_{10}$ and the $k$ $NS5$ branes become
$k$ $M5$ branes which are located at the points $z_1, \ldots, z_k$
on the torus $T^2$. The $N$ $D4$ branes are lifted to $N$ $M5$ branes
which wrap the whole of the $T^2$.

If the ALE singularity was of $D$ type then in addition to 
$NS5$ branes one finds orientifold plane on Type $\II$A side.

If the $NS5$ (or $M5$) branes are dislocated in the $789$ plane then
the corresponding ALE space is resolved. The parameters 
$\vec r_i - \vec r_j$ describing the relative positions
of the fivebranes in the $789$ plane are mapped
to the hyperkahler moduli of ALE space. 
The corresponding process in the field theory is described by
turning on the FI terms $\vec \zeta$. 

The plane $45$ transverse to the $D4$ branes has the meaning of
the Coulomb branch direction. Let us denote by $\phi = x^4 + i x^5$
the corresponding coordinate. Then the separation of the $D4$ branes
in the $\phi$
direction causes the $NS5$-branes to be frozen at the same point $\vec
r$ in the
$789$ plane and vice versa. Of course, this is the familiar picture of
the
transitions between the Coulomb and Higgs branch with or without FI
terms.

Now let us rotate some branes. Let $x^{7} + i x^{8} = x$ be another
holomorphic
coordinate. Consider tilting the $NS5$-branes in such a way that for
$i$'th brane its position in $78$ plane linearly depends on $\phi$:
\eqn\bps{x_i = \xi_i \phi_i}
This configuration of fivebranes preserves supersymmetry (one can
think of it as of the fivebrane ``wrapping'' a holomorphic
curve $\prod_i ( \zeta - \xi_i \phi ) = 0$) and leads to $\CN=1$ gauge
theory
on the world-volume of $D4$-branes. The tilting makes the scalars in
the vector multiplets of $\CN=2$ supersymmetry massive, since the $D4$
branes are no longer free to slide along the $NS5$ branes in the
$45$ directions. Another way of
seeing
this is to notice that the condition that FI term is proportional to
the
scalar in the vector multiplet is identical to the equation \mmp. 

Finally, let us consider what happens when we add D5-branes to the
stack of $N$ D3-branes on an orbifold singularity.  For simplicity we
will restrict our attention to an $A_{k-1}$ orbifold singularity which
has been resolved by FI terms: the geometry is a direct product of the
4-dimensional ALE space (dimensions 6789), the complex plane
(dimensions 45), and flat Minkowski space (dimensions 0123).  As
remarked above, there are $k$ 2-cycles which sum to zero in homology
and through which there are fluxes $\theta_i - \theta_{i-1}$ of the NS
B-field.  Consider wrapping a D5-brane around one of these cycles,
with its other dimensions in the directions 0123.  The term linear in
$B_{NS}$ in the Wess-Zumino part of the D5-brane action \jpTASI,
  $$i\mu_5 \int e^{2\pi\alpha' F + B_{NS}} C$$
 gives this wrapped D5-brane precisely $(\theta_i-\theta_{i-1})/2\pi$
of a D3-brane charge.  This is a special case of the phenomenon of
fractional branes and wrapped branes discussed in \dm\enh\ddg, only here we
are allowing arbitrary $\theta_j$ rather than taking $\theta_j = 2\pi
j/k$.  Upon T-dualizing, the D5-brane becomes an extra D4-brane
stretched between the $i-1$'st and $i$'th NS5-branes.\foot{We thank
A.~Karch for bringing to our attention the reference \kls, which includes
a similar discussion of this construction.}  This has the
effect of changing the gauge group: it was $SU(N) \times \ldots \times
SU(N)$, with $k$ factors of $SU(N)$; now the $i$'th gauge group
becomes $SU(N+1)$.  The supersymmetry is still ${\cal N}=2$, and the
hypermultiplets are still in bifundamental representations.  The
interpretation of $D5$ branes wrapped on a 2-cycle as modifying a
gauge theory by incrementing the rank of one gauge group was suggested
in \gkBaryon\ based on evidence from anomalous brane creation.  The
use of T-duality in a perturbative D-brane setting reinforces that
interpretation.

It would be nice to T-dualize back from brane realizations of gauge
theories to obtain the exact supergravity/string background which are
dual to them, similarly to the construction for $A_1$ case in
\mukhi. Unfortunately, at the moment it does not seem to be very
practical.

\newsec{Conclusions and conjectures}
\seclab\Conclusions

So far we described a class of complex threefolds which generalize
the ordinary conifold.  Our construction is most easily described in
the language of the gauge theory on the world-volume of the probe
$D3$-brane
placed at the singularity of the threefold. We start with the quiver
$\CN=2$ gauge
theory of the ADE type which corresponds to the manifold
which locally looks like $Y_{\Gamma, {\rm UV}} = \IC^2/{\Gamma} \times
\IC$. 
The manifold $Y_{\Gamma, {\rm UV}}$ is a cone over the base
$M^{5}_{\rm UV} = S^{5}/{\Gamma}$. When the large $N$ number of 
$D3$-branes are placed at the singularity they can no longer be
treated as  probes. Instead, they change the space-time geometry from
that of $\IR^{1,3} \times Y_{\Gamma, {\rm UV}}$ to 
$AdS_{5} \times M^{5}_{\rm UV}$ and there is a flux of RR five-form
field through
 $M^{5}_{\rm UV}$ which is equal to $N$. The properties of the string theory
propagating
in this background are believed to be reflected  in those
of the superconformal gauge theory which occurs at the origin of the
space
$Y_{\Gamma, {\rm UV}}$ considered as a Higgs branch of the gauge
theory on branes.

The $\CN=2$ superconformal  theory has a number of interesting
deformations.
It is known that it has exactly $r+1$ complex marginal deformations
corresponding to the
couplings of various gauge factors. Their space-time counterparts
are the space-time dilaton+axion $\tau$ and the fluxes
of the RR and NSNS $B$-fields through the collapsed two-cycles
which are fibered over the fixed circle in $S^{5}/{\Gamma}$ \lnv. 
The six-dimensional tensor multiplet which contains these fluxes 
also contains the
parameters of the deformations of the two-cycles themselves (three
parameters per cycle). These would correspond to the FI terms in the
gauge theory. 
The $\CN=2$ gauge theory deformed by the generic FI terms flows
to the trivial IR fixed point. The space-time interpretation of this
fact is that if one first resolves the orbifold $\IC^2/\Gamma$ into
a smooth space and then places the large number of threebranes at
the generic point of it then the near-horizon geometry will be
$AdS_{5} \times S^5$ as in the absence of any orbifold. 

There are two distinct claims one could make regarding D3-branes
located at the Calabi-Yau singularities we have described.  The first
and simplest is as follows: given a conifold singularity of a
particular ADE type, the low-energy theory of D3-branes located at
that singularity is the IR fixed point arising from a $\CN=2$ theory
deformed by giving masses to the $\CN=1$ chiral multiplets within the
$\CN=2$ vector multiplets, as described in section~\GaugeTheory\ and
\GenAllN.  The $\CN=2$ origin of the gauge theory begs the question in
what sense one can start with D3-branes at an ADE orbifold singularity
and ``flow'' to the conifold geometry.  It was argued in the $A_1$
case in \witkleb\ that there is no topological obstruction to the flow
(more specifically that a resolution of $S^5/\IZ_2$ has the same
topology as $T^{11}$).  We have taken one more step toward describing
such a flow by identifying the multiplet of $AdS_5$ supergravity which
includes the blowup modes and observing that from in $AdS_5$ some
fields in this multiplet have just the right tachyonic masses to
correspond to the scalar and fermion masses involved in deforming the
gauge theory.  The states in this multiplet arose in the analysis of
\gukov\ from the twisted sector localized at the circle on
$S^5/\Gamma$ fixed by the action of $\Gamma$.  In a nutshell the
second claim is that starting with D3-branes at an ADE orbifold
singularity, one can turn on fields which in the gauge theory are the
mass deformations and in the string theory are twisted sector modes,
and obtain a string theory background which tracks the RG flow which
takes the gauge theory from its UV fixed point (with $\CN=2$
supersymmetry to its IR fixed (with $\CN=1$).  

Assuming such a string background exists, what are its properties?
It should have the rotational and translational symmetries of
four-dimensional Minkowski space, and it should preserve four real
supercharges.\fixit{a background with nontrivial t-dependence is hard
for me to take seriously} Five-dimensional supergravity is a valid
description of the low-energy dynamics of both ends of the flow
(provided we take $N$ sufficiently large and include the matter
multiplets arising from the twisted sector of the orbifold), so it
seems likely that it is in fact a valid throughout the flow.  It is
not clear whether truncating the theory to a small number of
multiplets (as was done in effect in \ppz\ and \dz) is a controlled
approximation far from the fixed points.  The $AdS_5$ metric should be
recovered at either end of the flow (although with different radii,
related to the central charges as in \gEin), and in the full
ten-dimensional string theory description we expect to see the metric
smoothly approach the factorized form $AdS_5 \times M^{5}_{\rm UV}$ in
the ultraviolet and $AdS_5 \times M^{5}_{\rm IR}$ in the infrared.
$M^5_{\rm UV} = S^5/\Gamma$ as above, and $M^5_{\rm IR}$ is the base
of the cone described in section \BaseGeom.  The total space of the
cone is the Calabi-Yau manifold, whose complex structure we described
in the previous sections.  It is not clear to us what on this cone
should play the role of a radial coordinate, dual to scale in the RG
flow.

To put it in a single phrase, the two ends of the RG group flow
correspond to the two Einstein manifolds, $M^5_{\rm UV}$ and $M^5_{\rm
IR}$.

\newsec{Acknowledgements}

We would like to thank O.~Aharony, J.~Distler, M.~Douglas, S.~Gukov,
Y.~Oz, M.~Rozali
and
C.~Vafa for discussions.  The research of S.~G.{} and N.~N.{} was
supported by Harvard Society of Fellows, and also in part by NSF grant
number PHY-98-02709.  The research of N.~N.{} was also partially
supported by RFFI under grant 98-01-00327 and by grant 96-15-96455 for
scientific schools.  The research of S.~S. is supported by DOE grant
DE-FG02-92ER40704, by NSF CAREER award, by OJI award from DOE and by
Alfred P.~Sloan foundation.  S.~G.{} and N.~N.{} are grateful to Aspen
Center for Physics for hospitality during the initial phases of this
work. S.~G.{} also thanks the University of Chicago for hospitality
during the later phases of the work.
N.~N.{} wishes to thank Physics Department at Rutgers University
for hospitality at the final stages of preparing the manuscript.

\listrefs
\bye